# HCI considerations in Designing a Second Life Virtual Therapeutic Community for the Support & Treatment of People with Borderline Personality Disorder

Alice Good[1], Paul Gnanayutham[2], Arunasalam Sambhanthan[3], Vahid Panjganj[4]

[1]Department of Computing, University of Portsmouth, Buckingham Building, Lion Terrace, Portsmouth PO1 3HE, United Kingdom, Tel: +44-(0) 23-9284-6664, alice.good@port.ac.uk

[2]Department of Computing, University of Portsmouth, Buckingham Building, Lion Terrace, Portsmouth PO1 3HE, United Kingdom, Tel: +44-(0) 23-9284-6468, paul.gnanayutham@port.ac.uk

[3]Department of Computing, University of Portsmouth, Buckingham Building, Lion Terrace, Portsmouth PO1 3HE, United Kingdom, arunsambhanthan@gmail.com

[4]Department of Computing, University of Portsmouth, Buckingham Building, Lion Terrace, Portsmouth PO1 3HE, United Kingdom, vahid.panjganj@myport.ac.uk

**Abstract**

This paper aims to present the current position of the ongoing research into developing the requirements and acceptance of a virtual therapeutic community in Second Life, specifically for people with Borderline Personality Disorder (BPD). The research has identified this particular user group given that people with BPD often require high levels of support, which can result in emergency hospital admissions, in addition to the significant economic cost of treating people BPD in relation to other mental illnesses (NCCMH, 2009). The research is also intended to be used as framework for other mental health conditions. This work is a continuation of the research carried out in exploring the potential of virtual therapeutic communities based on existing models of therapeutic hospitals as well as virtual treatments and support, in treating people with BPD. An interdisciplinary approach to this research features collaboration from areas in HCI, forensic psychology and psychotherapy.

## 1. Introduction

The fourth version of Diagnosis & Statistical Manual of Mental Disorder classifies Borderline Personality Disorder as a mental disorder which is characterized by a lack of one's own identity, with rapid changes in mood, intense unstable interpersonal relationships, marked impulsively, instability in affect and instability in self image (DSM-IV; APA, 1994) People with Borderline Personality Disorder (BPD) represent 0.7% of the UK population (NICE, 2009) and are reported as more likely to seek psychiatric intervention than those with other psychiatric disorders (Rendu *et al.*, 2002, cited in NICE, 2009). Whilst the use of ICTs in providing support for people with mental health problems is certainly expanding, treatment for severe mental illness however does not often consider the potential that social networking can have in both reducing the impact of loneliness (Perese & Wolf, 2005), and the sense of 'feeling alone' (Neal & McKenzie, 2010). Certainly, there is a high need for support for these people yet research suggests that some mental health professionals working with people with BPD are not aware of the provision of ICT enabled support, including social networking and static information sites. (Good *et al.*, 2011).

The focus of this paper is to present the position of this ongoing research into designing a virtual support system for people with BPD. It is anticipated that this research can be used as a framework for other mental health conditions. The justification for the focus on people with BPD specifically is two-fold. One, these people are more likely to seek psychiatric intervention and secondly, incurring primary care costs at almost twice the amount as patients with other mental illnesses (Rendu *et al.*, 2002, cited in NICE, 2009). The research is very much user driven and aims to encompass HCI principles both in the requirements and design of the proposed system. This paper will look at literature relating to HCI considerations for the design and evaluation of Second Life and other virtual world environments. It will then present the justification of the need for additional support for people with BPD. Furthermore, it presents the basic requirements in terms of usability, accessibility and security.

## 2. HCI considerations for a Virtual Support System Hosted in Second life

This section looks at the characteristics of second life as a proposed environment for a virtual support system. Furthermore, it reviews usability and accessibility considerations as well as the need for specific evaluation frameworks.

### 2.1. *Second Life and usage*
Some literatures define Second life using its characteristics. 3D environment, user avatars, moving around the visual world and interaction with other objects, are the main functionalities that have been used to describe the Second Life (Varvello, Picconi, Diot, & Biersack, 2008). Among those, interaction between avatars has established such opportunity for socializing, which in result many authors define Second Life as a virtual world for meetings and joint activities (Tay, 2010; Lucia *et al.*, 2008). However these definitions hardly distinguish Second Life from any other virtual environment. Accordingly, Menchaca *et al.,* (2005) reminds us that Second Life is basically a CVE (*3D* Collaborative Virtual Environment), which puts collaboration as the main purpose of the avatar. Kumar *et al.,* (2008) go even further and separate SL from other CVEs by pointing out attributes such as seamless persistent world, User-generated content and massive and dynamic content. Finally Veerapen (2011, p. 261), clearly defines Second Life, as "*... a persistent, non gaming, collaboratively user-produced, mediated, virtual world, which permits multiple users, represented through avatars, to interact synchronously with one another and the environment*".

   Collaboration is the main focus of any CVE including SL (Second Life). Consequently, it is being used widely for victual conferences, meetings (Erickson *et al.*, 2011; Berg van den, 2008) and indeed Erickson *et al.* (2011, p. 4), in their survey show that 66.2% of participants in one of IBMs virtual meetings consider it as a good experience. However, SL's potential can be used for other purposes such as educational (De Lucia *et al.,* 2009), which is due to its alignment with the concept of experiential learning (e.g.. better demonstrations of complex scientific concepts, rich media content for learning, greater learning autonomy etc.) (Perera *et al.,* 2010). Moreover, SL's usage covers areas including Clinical Psychology (Gorini *et al.,* 2008), retail sales, virtual tourism, marketing and disaster recovery training (Kumar, *et al.,* 2008) and also support groups.

   A recent survey on healthcare related activities using Second Life shows that patient education and awareness building as the major health related activity undertaken through second life. The second largest group of sites was that of support groups (Beard *et al.,* 2009). Research by Norris (2009) looked at the growth of healthcare support groups in virtual worlds and reported that mental health groups featured the largest number of members at 32% of the total users of Second Life. In terms of categories of groups, 15% of the health support groups in Second Life were dedicated to mental health. Second Life as a social networking medium then holds some potential in facilitating support and information sharing for people with BPD. Furthermore, there are some examples of Second Life facilitating therapy. It is reported that virtual worlds can be effective in confronting phobias and addictions as well as offering potential in experimenting with new behaviours and means of expression. Research shows that Second Life has been a usual tool in facilitating exposure treatment for anxiety and behavioural activation for depression (Newman *et al.,* 2011).

2.2. *HCI, Second Life & Evaluation*

Studies including one by Bessière *et al.* (2009), have mentioned the fact that infarctions are the building pieces of any virtual environment. However in Collaborative Virtual Environments such as Second Life, this concept goes even further since the user not only interacts with the application, its avatar is interacting with objects and other people in the environment as well. Analysis of interactions involved in CVEs and the HCI factors that differentiate it from other VEs was first carried out by Steed & Tromp (1998) and was later supported by Veerapen (2011, p. 261).

Essentially, the main aim of HCI evaluation and improvement from Bessière's *et al.* (2009) point of view is to attract potential participants who prefer face-to-face interaction in collaborative works. The viewpoint of these participants is that nothing will ever replace the face-to-face interaction. However, advantages of HCI evaluation in this field are not limited to that. Positive impacts of HCI in CVEs from two different perspectives have been demonstrated by studies. Berg van den (2008) defines a framework to determine Second Life's suitability. It later shows that even though using Second Life has lots of advantages, for example for a company like IBM, lack of the suitability can prevent the company from spending valuable resources. In another study, Looi & See (2010), deploy HCI evaluation methods in Second Life in order to effectively improve students' activities in Second Life. They discuss that the first step that achievement is to "...*examine the Human Computer Interaction facets*..." (Looi & See, 2010, p. 2). They explain how HCI methods in supporting user engagement improve the quality of educational experience in Second Life. Nonetheless, they argue that HCI methods only ease the problems faced in virtual learning and never provide any direct solution for existing challenges.

2.3. *HCI evaluation frameworks for Second Life*

Evaluating usability is essential for the acceptance and success of any software. Of course collaborative virtual environments (CVEs) are not exceptional and there have been evaluation frameworks available before Second life was even launched. Steed & Tromp demonstrated a HCI evaluation framework experimentally in 1998. They used two prototype CVE applications, which were developed by the COVEN project. Steed & Tromp (1998), discuss that methodological constraints specific to CVE evaluation, plus absence of an existing dedicated CVE evaluation methodology, move them toward the design of a framework for the usability evaluation. According to their findings, those constraints relevant to nature of CVEs arise from the fact that understanding human behavioural needs is essential for development of CVE components. Consequently, for evaluation "*...one has to strike a balance between the concerns of usability engineering and scientific enquiry frameworks*" (Steed & Tromp, 1998, p. 3). The framework had three main threads of work.

1. Monitoring the performance of initial use cases in order to detect basic errors. Using four independent inspectors, issues were captured and classified into three categories of System problems, Interface problems and Application specific problems.
2. Observational evaluation of users while performing tasks in order to understand human behavioral concepts in CVE.
3. "Isolated auxiliary case-controlled experiments" which in their view emulates aspects of the CVE concept.

The Affordance-based evaluation framework for CVEs was introduced by Turner & Turner (2002), who believe usability in CVEs involves both user-UI and virtual collaborative interactions. They developed a three-layer model of affordances to be used as an evaluation framework. This model comprises Usability, embodiment and purposive. They describe level one (usability) as basic interactions, such as using mouse for moving in the environment. In level two which is "Affordances supporting user tasks" they take tasks and subtasks into consideration. They propose that the accomplishment of users in CVE's is basically focussed upon the concept of collaboration, which is highly dependent on embodiment. Finally in the third level, the focus is on cultural affordance. Turner & Turner (2002), introduce it as a concept to cover specific tasks carried out by specific users for specific reasons. They emphasise the fact that level three can be measured only by member of the group who create the task, as evaluation in this level is so use-focussed nature and highly contextual. They later provide examples to explain how the third layer distinguishes different usages of the CVE.

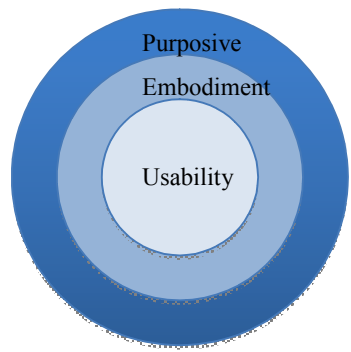

**Figure: 1** *Affordance-based evaluation framework layers*

Turner & Turner's (2002) work produced a platform, which helped Berg van den (2008) to develop his own evaluation framework specifically for Second Life. This framework was used to evaluate the suitability of Second Life based on three main components. According to Berg van den (2008) the first component is meant to describe the interaction capability and communication functionality followed by the second component, which identifies supporting interaction capabilities (e.g. navigation). Finally, the last component gathers the requirements for a specific application of second life. Using the framework Berg van den (2008) compares the requirements against basic (component one) and supportive capabilities (component two) to determine usability of Second Life toward a specific application.

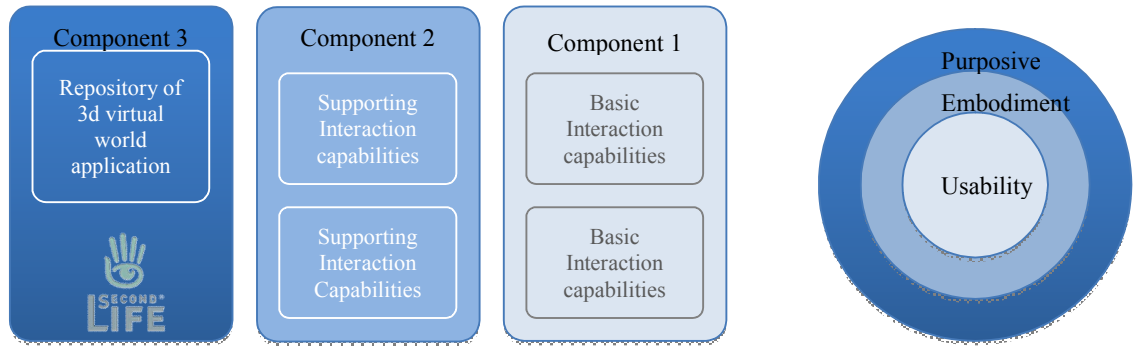

**Figure 2:** *From Affordance-based evaluation framework (right) to Berg van den's framework (left)*

Berg van den (2008), keeps the framework flexible enough (using component three) to be used for any type of application in the Second Life, De Lucia *et al.* (2009) however, develop an evaluation framework, which considers learning aspect of virtual worlds and focuses on educational purposes of the second Life. According to them to evaluate effectiveness of a CVE following factors are vital: presence, awareness, communication, and the perception of belonging to a learning community.

Presence arises from the feeling of being part of the world. De Lucia *et al.* (2009) believe that "*...presence and learning are strongly related;*" (p. 220) and developing this factor can significantly improve the learning by making the experience more meaningful. Awareness is all about knowledge of the user from the environment, which effects the time it takes for them to participate. Awareness must help the user on locating themselves and provide users with answers for questions such as "Who is there", as well as ''what is going on" (De Lucia *et al.,* 2009). By emphasising on communication, as an important factor in evaluation, they suggest that verbal communication won't be enough on its own and a composition of non-verbal communication with visual body language will improve the effectiveness. At last but not least, "Belonging to a community" is what they describe as distinguishing between CVE as a social environment and CVE as a Learning environment.

The ADA 1990 is part of the US Public Law that prohibits employers from all sectors and agencies from discriminating individuals with disabilities amongst other equally qualified individuals.

*2.4 Accessibility and Second Life*

With the growth of Second Life in many areas of society, accessibility is an important consideration. Whilst the aforementioned literature addresses usability, there are no frameworks there is no provision for the support of accessibility evaluation. The percentage of Internet users exhibiting a disability or impairment is approximately 14% (Fonseca *et al.,* 2009, p. 1155). Despite the fact that Second Life has the potential to improve the quality of life for disabled people (Epstein, 2008), Vickers *et al.* (2008) believe that accessibility issues such as communication difficulties are due to designed interfaces, and online virtual environments are not intended to aid disabled users.

According to White *et al.* (2008), Second Life is inaccessible for the majority of visually impaired users. A study by Oktay & Folmer (2010), which traces the improvement of accessibility in Second Life by development of a screen reader accessible interface, failed to make virtual world accessible to people with visual impairments due to factors such as lack of Meta data for many virtual objects in Second Life. Therefore, White *et al.* (2008) promote the evaluation of even the very basic interactions using both disabled (blind and VI) users and researchers in order to identify the source of problems. White *et al.*(2008), evaluate Second Life based on two main principles, semi-structured interviews with blind and visually impaired people and accessibility evaluation against four scenarios. In interviews, disabled users share their personal experiences with accessibility issues in both real and virtual environments. The scenarios are basic activities, which monitoring user engaging these tasks helps to identify accessibility issues within them. The scenarios are Content Creation (as one of the main features of SL), Trade, Spectacle (e.g. Avatar customization) and Communication.

Vickers's *et al.* (2008) HCI evaluation framework in comparison with White *et al.* (2008) chooses the use case scenarios at the basic level. They evaluate the HCI based on main control and manipulation areas available in CVEs, such as locomotion and camera movement; object manipulation; application control and communication. They then evaluate control requirements for those use cases against accessibility levels. For instance, using arrow buttons located on semi-transparent overlays are required for Locomotion and camera movement. Object manipulation, which is only controllable using mouse. Application control that is accessible via mouse and shortcut keys, and finally communication, achieved through text generation or voice via a microphone (Vickers, Bates, & Istance, 2008). By summarizing those use cases according to control source and their task domain. The framework allows the analyst to measure the accessibility based on control source over variety of tasks and provide recommendations in order to improve the HCI for people with disabilities. For instance Vickers *et al.* (2008) recommend Gaze tracking as an effective replacement for some current sources control to help users with high levels of paralysis. Unlike the White's *et al.* (2008) work, which is focused on visually, impaired users, this framework support accessibility evaluation by considering different disabilities.

Table 1: *Summary of task domains and control sources used by* (Vickers, Bates, & Istance, 2008)

|  | Control source | | |
| --- | --- | --- | --- |
| **Task domain** | Mouse | Keyboard | Speech |
| Locomotion and camera movement | x | x | x |
| Object manipulation | x | x | x |
| Application control | x | Partial | x |
| Communication | x | x | x |

As with any interface, usability and accessibility are fundamental in ensuring acceptance and positive user experience. Certainly, traditional empirical user testing will help in identifying potential issues. The aforementioned literature however highlights specific methods in addressing HCI, including evaluation frameworks that are focussed upon Second Life or other virtual world environments.

## 3. Justification and Current Position of Research

### 3.1 Borderline Personality Disorder in Perspective

BPD is a debilitating disorder, causing great distress to those that exhibit the disorder and close friends and family of the person. People with this disorder exhibit a range of debilitating and self destructive behaviours including: depression, poor social skills and instable relationships; chemical dependency; eating disorders and suicide attempts (DSM-IV; APA, 1994). Of greater concern is the fact that approximately 60-70% people with BPD are reported to commit suicide, with approximately 10% actually being successful (Oldham, 2006, cited in NICE, 2009). In a study conducted by Moran *et al.* (2000), also cited in NCCMH (2009), it is reported that 4-6 % of patients receiving primary care have been diagnosed with BPD. Support and understanding of these individuals is then paramount. Treatment for people with BPD has traditionally included both pharmacological and psychological intervention, the latter including: Cognitive Behaviour Therapy (CBT); Dialectical Behaviour Therapy (DBT) (Lineham *et al.,* 1999) and Therapeutic Communities (TCs).

### 3.2 Therapeutic Communities as an effective treatment model for BPD

It is the concept of therapeutic communities that forms the basis of this research, which initially originated from the treatment of veterans suffering with PTSD from the Second World War (Main, 1946). The recently closed Henderson Hospital that emerged in the 1950s is an example of a TC featuring four specific themes: democratisation, permissiveness, reality confrontation and communalism (NICE, 2009). TCs essentially incorporate a democratic ethos and are run with less formal roles from the professionals, with both members and professionals having an equal say in the organizations and management (NICE, 2009). Research into the effectiveness of TCs, show that they are a viable and productive method for treating BPD (Norton, K. & Hinshelwood, R. D.,1996; Campling, P.,1999) and indeed the *only* recommended residential treatment (NICE, 2009). Research has been shown to validate the effectiveness of TCs in both the reduction in symptoms (Dolan et al, 1997) and in primary care costs (Davies, 2003) one year following discharge. However, whilst they are shown to be a useful means of treatment, a change in primary care funding, now targeted towards outreach teams, has resulted in the closure of some TCs, with others now under threat. Indeed, this has led to fear and uncertainty for those that have experienced traumatized childhoods with, not only the negative connotations associated with the label BPD but also the plain knowledge that a viable and effective treatment is now no longer an option (Thomas, 2011). Furthermore, with attitudes from some mental health professionals that BPD is 'untreatable' and a lack of awareness in specific BPD online support systems (Good et al, 2011), many sufferers of BPD remain lacking in necessary treatment and support. It is precisely the demise of TCs that has provided the motivation for a virtual TC.

### 3.3 Current position of Research

This research is looking at a way of developing a virtual therapeutic community in Second Life as a support system for people with BPD. This type of support system would emulate many of the features prominent within a therapeutic community. Certainly for this proposal to be considered viable and accepted by both clients and professionals, it is paramount for user centered approach to all stages of the project. The project is in its initial stage. Research has been carried in understanding mental health professional's awareness of online support for people with BPD (Good *et al.*, 2011). A client perspective of the availability and need for support, both online and face to face is currently being conducted. A comparison of client versus professional perspective will be made. Current research is also looking at the level of support on social networking sites for people on BPD. Further studies looking at the specific requirements will be carried with end users very much involved at all stages.

## 4. Framework of Requirements

The requirements are classified into specific categories in the table below. The usability of second life based therapeutic communities has been emphasized by Turner (2002) and Berg Van Dan (2008). The specific usability requirements of a second life CVE are derived from the literature and contextualized to the proposed system. The research of White et al (2008), Octay & Folmer (2010) and Vikers *et al*. (2008) clearly indicates the requirements for accessible design of virtual worlds for people with impairments.

**Table 2**: *Aspects of requirements based upon the literature*

| Aspect | Specification | Rationale |
|---|---|---|
| Usability | - Basic Interactions: Use of mouse<br>- Affordances supporting user tasks: Clear tasks and subtasks<br>- Cultural affordance: A user specific task analysis | Essential framework elements proposed by Turner (2002) for usability evaluation in CVEs. |
| | - Interaction Capability: Avatar Customization<br>- Communication Functionality: Chat Forums<br>- Interaction capability: Navigation | Proposed by Berg Van Dan (2008) specifically for second life. The framework could be further enhanced |
| Accessibility | - Accessible interface<br>- Sound facility for visually impaired users | White *et al*. (2008), Octay & Folmer (2010) and Vikers *et al*. (2008) emphasise the importance of including |

|  | • Use of arrow buttons located on semi transparent overlays | accessibility requirements in the second life based virtual community systems. |
|---|---|---|
| Security | • Authentication mechanism<br>• Privacy enhanced discussion forums<br>• Authorization for accessing discussion forums<br>• Anonymity preference to hide the identity of client to other clients | Basic security requirements are essential to ensure anonymity, data protection and privacy. |

## 5. Conclusions

The premise of this research is to develop a virtual 3D world TC in Second Life, specifically for people with BPD. Such an environment would emulate some of the core principles of TCs, as they have been shown to be effective in treating BPD. Second Life would certainly enable an enhanced sense of immersion and presence as opposed to static methods of communication and interaction. Such a proposal will undoubtedly require the collaboration of specialised professionals and the co-operation of outreach services, as well as the input of end users. In terms of the design process, an end-user centred approach would be applied where clients presenting a diagnosis of BPD contribute to the viability of this work, along with professionals specializing in the treatment of this condition. A triangulation approach will be applied in the requirements stage, using a focus group and interviews to validate the results and to provide a deeper understanding.

This paper has looked at the issue of BPD and how sufferers require high levels of support, particularly when in crisis. Some of these people make use of various online forums. Second Life has been shown to be very popular in facilitating mental health support groups, yet these are currently discussion based and as yet, there are none specifically dedicated to sufferers of BPD. The paper has explored the importance of adhering to specific evaluation frameworks when designing virtual worlds, focussing on HCI aspects such as usability and accessibility. The current position of the research project is also highlighted.